\documentclass[journal=jpclcd,manuscript=article]{achemso}
\usepackage{chemformula} 
\usepackage[T1]{fontenc} 

\usepackage{graphicx}  	
\usepackage{bm}        	
\usepackage{amssymb}   	
\usepackage{amsmath}
\usepackage{ulem}

\newcommand{\li}[1] {\textcolor{red}{#1}}

\usepackage{xcolor} 
\usepackage{soul} 

\author{Nicole M. James}
\affiliation{James Franck Institute, The University of Chicago, Chicago, Illinois 60637, USA}
\alsoaffiliation{Department of Chemistry, The University of Chicago, Chicago, Illinois 60637, USA}
\alsoaffiliation{These authors contributed equally to this work.}

\author{Chiao-Peng Hsu}
\affiliation{Laboratory for Interfaces, Soft Matter and Assembly, Department of Materials, ETH Zurich, Zurich, Switzerland}
\alsoaffiliation{Laboratory for Surface Science and Technology, Department of Materials, ETH Zurich, Zurich, Switzerland}
\alsoaffiliation{These authors contributed equally to this work.}

\author{Nicholas D. Spencer} 
\affiliation{Laboratory for Surface Science and Technology, Department of Materials, ETH Zurich, Zurich, Switzerland}

\author{Heinrich M. Jaeger} 
\affiliation{James Franck Institute, The University of Chicago, Chicago, Illinois 60637, USA}
\alsoaffiliation{Department of Physics, The University of Chicago, Chicago, Illinois 60637, USA}

\author{Lucio Isa} 
\affiliation{Laboratory for Interfaces, Soft Matter and Assembly, Department of Materials, ETH Zurich, Zurich, Switzerland}
\email{lucio.isa@mat.ethz.ch} 

\title{Tuning Interparticle Hydrogen Bonding in Shear-Jamming Suspensions: Kinetic Effects and Consequences for Tribology and Rheology}

\begin{document}
	
	\begin{tocentry}
		\begin{center}
			\includegraphics[height=5cm]{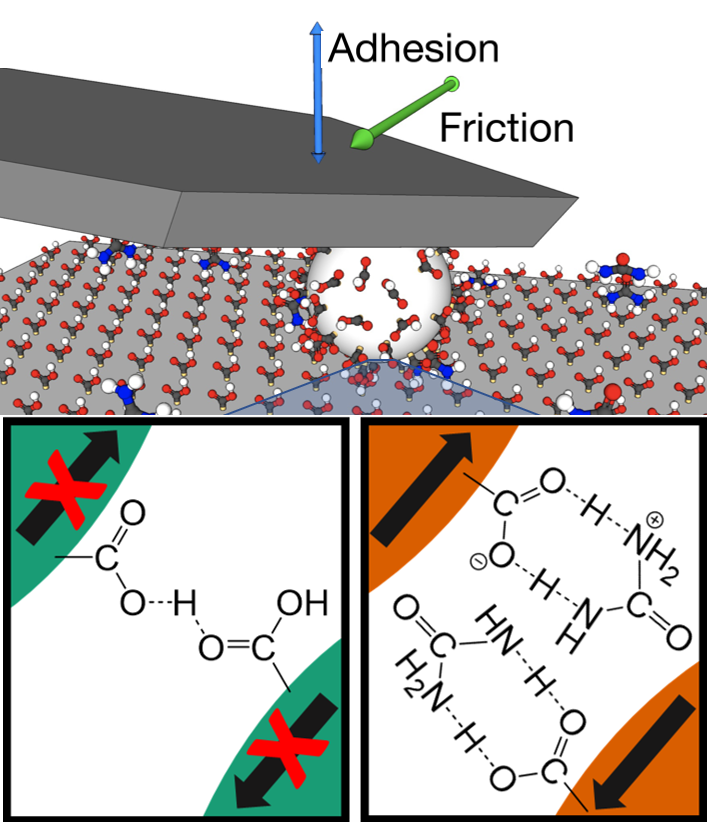}
		\end{center}
	\end{tocentry}
	
	
	
	\begin{abstract}
		
		The shear-jamming of dense suspensions can be strongly affected by molecular-scale interactions between particles, e.g. by chemically controlling their propensity for hydrogen bonding. However, hydrogen bonding not only enhances interparticle friction, a critical parameter for shear jamming, but also introduces (reversible) adhesion, whose interplay with friction in shear-jamming systems has so far remained unclear. Here, we present atomic force microscopy studies to assess interparticle adhesion, its relationship to friction, and how these attributes are influenced by urea, a molecule that interferes with hydrogen bonding. We characterize the kinetics of this process with nuclear magnetic resonance, relating it to the time dependence of the macroscopic flow behavior with rheological measurements. We find that time-dependent urea sorption reduces friction and adhesion, causing a shift in the shear-jamming onset. These results extend our mechanistic understanding of chemical effects on the nature of shear jamming, promising new avenues for fundamental studies and applications alike.
	\end{abstract}
	
	
	Dense suspensions of colloidal or granular particles in a fluid can undergo striking non-Newtonian behavior. Not only can the suspension's viscosity increase by orders of magnitude at a critical shear rate, a phenomenon known as discontinuous shear thickening (DST),\cite{Barnes1989,Brown2010} but the suspension may even solidify under shear, \textit{i.e.}, exhibit what is called shear jamming (SJ).\cite{Smith2010, Peters2016}
	Recently, it was shown that DST and SJ depend on interparticle friction.\cite{Fernandez2013,Seto2013}
	While suspended particles can slide past each other at low shear, beyond a critical shear stress the hydrodynamic lubrication layers between particles break down and the resulting frictional contacts generate transient (DST) or stable (SJ) networks of force chains.\cite{Mari2014}
	Increasing the effective friction coefficient characterizing these contacts is predicted to increase the range of particle packing fractions over which DST and SJ are observable.  
	From an experimental standpoint, interparticle friction can be controlled by coating the particle surface with polymer layers of varying lubricity \cite{Vyas_Switching_2008, Deinaite_Friction_2010, Fernandez2013} or by modifying the surface topography, \textit{e.g.} by controlling surface roughness.\cite{Lootens_Dilatant_2005,  Hsiao_Rheological_2017, Hsu2018}
	Friction can also be controlled by tuning the chemical interactions between particles. In particular, in suspensions of colloidal poly(methylmethacrylate)/itaconic acid (PMMA/ITA) particles, surface carboxylic acid groups contribute to interparticle hydrogen bonding, which leads to enhanced frictional interactions. In the presence of urea, which interferes with hydrogen bonding, friction is reduced.\cite{James2018} However, hydrogen bonding also introduces a short-ranged adhesion, or "sticking" force. This force differs from irreversible, longer-ranged attractive particle-particle interactions, which typically lead to shear-induced flocculation and the development of a yield stress\cite{Zaccone_2010}, in that it is activated only when the lubrication layer is broken and is fully reversible when the local shear stress falls below a critical level. Such stress-dependent, reversible sticking is not accounted for in currently available models, and the manner in which it affects the inter-particle friction coefficient has remained an open question. 
	
	In this Letter, we address this by conducting liquid-cell, colloidal-probe atomic force microscopy (AFM) studies to measure friction and adhesion between PMMA/ITA colloids (diameter $d$ $\approx$ 0.8 $\mu$m), as a function of the solvent urea concentration, and connect these findings to both the chemical microstructure and the macroscopic flow behavior by means of NMR and and rheological studies, respectively. 
	

	\begin{figure}
		\centering
		\includegraphics[width=0.5\textwidth]{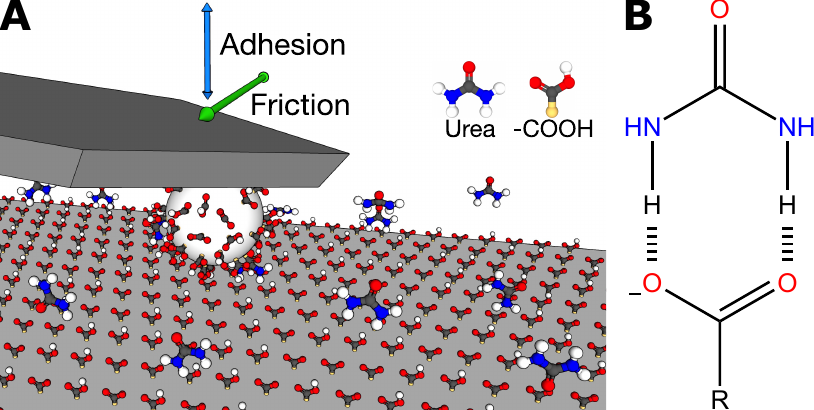}
		\caption{A. Schematic of a PMMA/ITA colloidal probe on a carboxylate-functionalized substrate in the presence of urea. The blue arrow indicates the cantilever movement while performing the adhesion measurements. The green arrow indicates the probe's scanning direction while performing the friction measurements. B. Representation of a possible coordination structure between urea and a carboxyl group. The dashed bonds represent hydrogen bonding interactions.}
		\label{Fig1}
	\end{figure}
	
	A schematic of the AFM experimental setup is shown in Figure \ref{Fig1}A. The PMMA/ITA colloidal probes used for friction and adhesion measurements are fabricated by attaching a PMMA/ITA particle to the end of a tip-less AFM cantilever (NSC36/Tipless, MikroMasch, Estonia) with a home-built micro-manipulator coupled to a microscope (BX 41, Olympus microscope, Japan). A small amount of epoxy glue is picked up with a sharpened tungsten wire (Wire.Co.UK, UK) and spread on the end of the tip-less AFM cantilever using the micro-manipulator. After that, a PMMA/ITA particle is picked by another sharpened tungsten wire and is affixed precisely over the epoxy glue.
	
	The substrate is a carboxylate-functionalized glass slide (AutoMate Scientific Inc.), selected to mimic the surface chemistry of the particle in order to provide a chemically symmetric contact. 
	AFM measurements are conducted with the particle and substrate immersed in the same suspending solvent used for the rheological measurements, 69\% aqueous glycerol (v/v), with urea concentration varied from 0-6 M. 
	
	As previously mentioned, urea interferes with the formation of hydrogen bonds between the scanning probe and the substrate, possibly following the scheme shown in Figure \ref{Fig1}B. \cite{Hay2005,Hughes1997} We hereby study how this bond formation affects adhesion and friction between carboxylated surfaces.
	
	\begin{figure}
		\centering
		\includegraphics[width=0.45\textwidth]{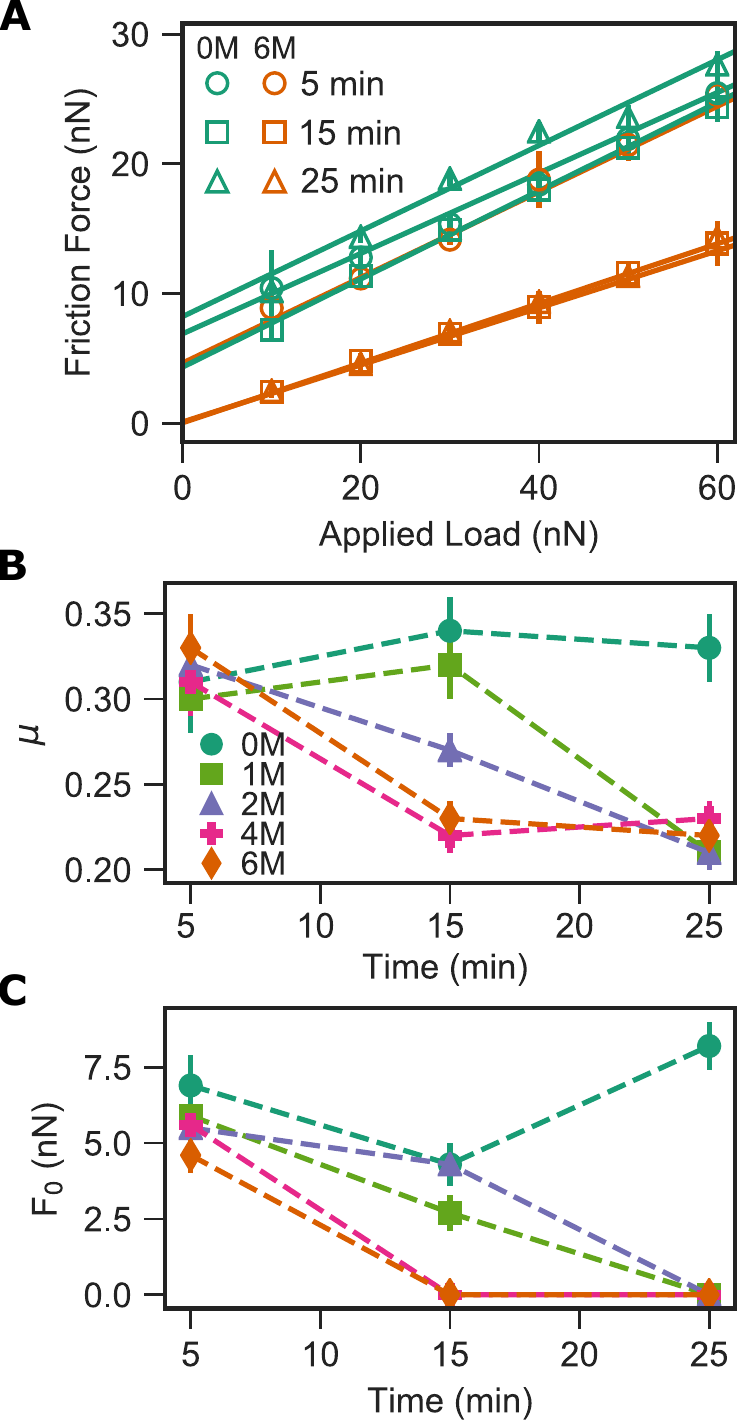}
		\caption{A. Friction measurements in 69\% (v/v) glycerol/water at 0 M and 6 M urea concentration. Error bars represent the standard deviations across the scan area (26 data points). The legend indicates how long the particle on the cantilever has been immersed in the solvent. B. Evolution of the friction coefficients over time, for various concentrations of urea. Error bars represent the uncertainties in the calculated slope based on Equation \ref{eqn1}.  C. Extrapolated force at 0 nN load, as a function of time for various concentrations of urea. Error bars represent the uncertainties in the calculated intercept based on Equation \ref{eqn1}.} 
		\label{Fig2}
	\end{figure}
	
	The friction measurements are carried out using lateral force microscopy by scanning the substrate in the direction orthogonal to the cantilever axis. The scan area and scan rate are fixed at 2 $ \mu $m $ \times $ 400 nm (128 px $ \times $ 26 px) and 0.5 Hz, respectively. The corresponding scanning speed is 1 $\mu$m/s, in order to avoid that viscous drag had any impact on the lateral deflection of the cantilever. The friction measurements are performed 5, 15, and 25 minutes after the system was immersed in 69\% glycerol/water (v/v), with urea concentration being varied from 0-6 M. At every scanning area, the friction loops are recorded at different applied loads $L$ from 10 nN to 60 nN.
	
	The friction forces as a function of $L$, with and without urea present, are shown in Figure \ref{Fig2}A for different waiting times prior to scanning. From the data, we see that the 0 M urea experiments lead to overall higher frictional forces compared to the 6M case and show no time dependence. Conversely, the 6M sample shows a history dependence. At short timescales ($<$ 5 min), the 6 M urea system yields similar results to the 0 M urea case, however, as time proceeds, the frictional force decreases, stabilizing after 15 minutes of exposure to the solvent. 
	The relation between friction force, $F_{\text{friction}}$, and applied load $L$ is described by a modified version of Amontons' Law\cite{Derjaguin_Molecuar_1934} as: 
	
	\begin{equation}
	F_{\text{friction}} = \mu\cdot(L_{0}+L) = F_{0}+\mu\cdot L,
	\label{eqn1}
	\end{equation}
	
	where a constant internal load $ L_{0} $ is added to the applied load $L$ to account for the intermolecular adhesive forces. $ F_{0} $ represents the friction force at 0 applied load for adhesive surfaces. The friction coefficient, $ \mu $, is defined as the slope of this equation, $ dF_{\text{friction}}/dL = \mu $. 
	
	The  friction coefficients as a function of time and urea concentration are shown in Figure \ref{Fig2}B. From these data it is clear that, in the absence of urea, the friction coefficient is stable over time, with a value of 0.33 $\pm$ 0.02. In the presence of urea, the friction coefficient reduces to 0.22 $\pm$ 0.01, a significant reduction of 34\%. This was previously reported to increase the particle volume concentration at which shear jamming is possible by $\sim$ 1\%\cite{James2018}, and it is in line with simulations that found a $\sim$ 55\% change in $\mu$ to result in a $\sim$ 3\% shift in the minimum particle volume concentration required for shear jamming.\cite{Fernandez2013}
	The speed with which the system equilibrates to this reduced $\mu$ value is highly dependent on urea concentration: the 1 M urea system takes between 15-25 minutes to equilibrate, whereas the 6 M urea system takes between 5-15 minutes. 
	
	In addition to the change in friction coefficient associated with urea concentration, we also find a marked difference in adhesion, as implied by the 
	non-zero value of the extrapolated friction force at 0 nN applied load,  $F_{0}$.\cite{Noy_Chemical_1995}
	In Figure \ref{Fig2}C, we show that $F_{0}$ bears the same urea- and time-dependence as that shown for $\mu$ in Figure \ref{Fig2}B. 
	This is consistent with the nature of the hydrogen bonding, which contributes an attractive interaction between surfaces that manifests as both resistance to sliding due to the constraints of bond lengths and bending, as well as interparticle adhesion. 
	The dependence of this timescale on the urea concentration suggests a kinetic process influencing this transition. 
	
	\begin{figure*}
		\centering
		\includegraphics[width=0.9\textwidth]{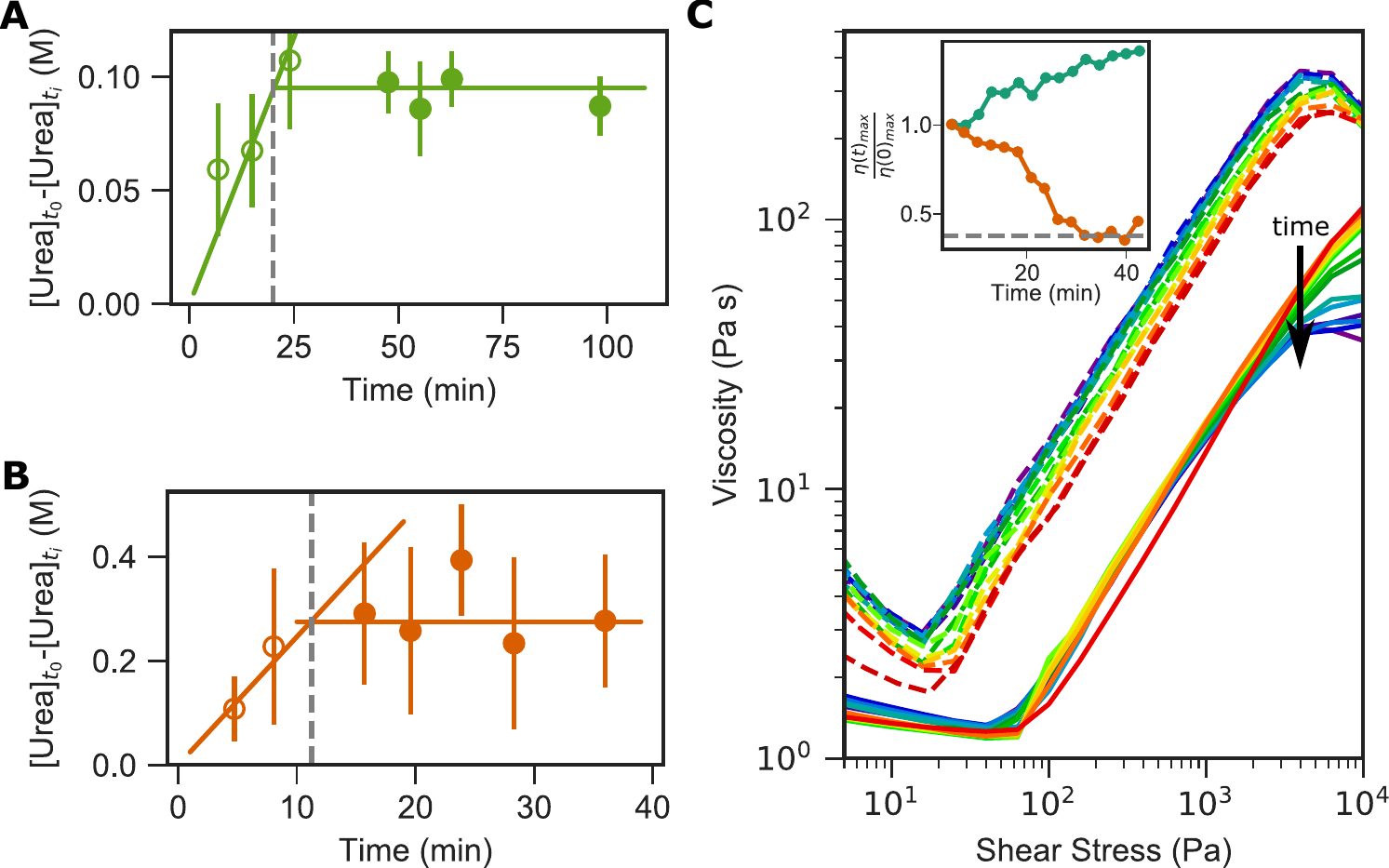}
		\caption{A. Equilibration timescales estimated by monitoring the magnitude of the change in urea concentration in the suspending solvent (69\% glycerol), as a function of particle immersion time. The initial urea concentration \sout{\li{(}}[Urea]$_{t_{0}}$ \sout{\li{)}} is 1 M. The open circles show the initial sorption period, where the concentration of urea is changing as urea molecules sorb to the particle surface; the closed circles show the steady state where the particle surface is saturated. The intersection of these two regimes suggests a timescale of $\sim$ 20 minutes (indicated by the gray dashed line) for the sorption process of dilute particles in 1 M urea. B. Equilibration timescales of dilute particles in 6 M urea. The intersection of the initial and steady state regimes suggests a timescale of $\sim$ 11.25 minutes (indicated by the gray dashed line). C. Rheology data as a function of time for 56\% PMMA/ITA microspheres in 69\% glycerol with 0M urea (dashed lines) and 6M urea (solid lines). Flow curves are taken sequentially over time ($\sim$2.5 minutes per flow curve); time increases from red to purple. Inset: the maximum viscosity at the peak of shear thickening, normalized by the initial (red) maximum viscosity, as a function of time, for 69 \% glycerol with 0 M urea (teal) and 6 M urea (orange).}
		\label{Fig3}
	\end{figure*}
	
	We directly connect this time-dependent change in friction and adhesion to the kinetics of urea sorption on carboxylated surfaces by quantifying the change in urea concentration of the suspending solvent over time. 
	From an initial exposure to urea at $t=t_0$, the concentration of urea in the suspending solvent decreases as it binds to the particle surface. 
	When a dynamic equilibrium is reached, the urea concentration in the suspending solvent stabilizes over time. 
	We measure this by exposing $\sim$ 1 g of PMMA/ITA particles to $\sim$ 5 mL of suspending solvent, under vigorous stirring. 
	At various timepoints, we withdraw $\sim$ 0.1 mL of suspending solvent and filter out the PMMA/ITA particles using a sterile, polyvinylidene fluoride syringe filter with a pore size of 0.22 $\mu$m (Fisher). 
	Approximately 0.03 g of suspending solvent is recovered, which is then diluted with 0.4 mL of deuterated dimethyl sulfoxide (DMSO-d$_6$), and 20 $\mu$L of 1,2-dichlorobenzene is added as an internal standard. 
	Urea is quantified by $^1$H NMR (number of acquisitions $=16$, relaxation delay $=20$ sec). 
	These results are shown in Figure \ref{Fig3} for 1 M urea (Figure \ref{Fig3}A) and 6 M urea (Figure \ref{Fig3}B). 
	Error bars indicate propagated uncertainties.
	We identify the system as equilibrated when the the change in urea concentration, $\text{[urea]}_{t_{0}}-\text{[urea]}_{t_{i}}$ has stabilized (indicated by filled circles). 
	Time points before this point indicate states where urea continues to sorb onto the particle surface (indicated by open circles). 
	By estimating the intersection of these regimes we identify an estimated timescale associated with the equilibration of the sorption process. 
	This timescale is $\sim$ 20 minutes for 1 M urea, and $\sim$ 11.25 minutes for 6 M urea. 
	These values are in good agreement with the timescales for the stabilization of $\mu$ and $F_{0}$, identified to be between  15-25 minutes and 5-15 minutes, respectively. 
	Thus, we conclude that the time-dependent evolution of the friction coefficient and the adhesion in the system is due to the sorption kinetics of urea on the PMMA/ITA particle surface. 
	
	We connect this with the macroscopic rheological properties of the suspension by rapidly measuring sequential flow profiles with and without urea present. 
	These results are shown in Figure \ref{Fig3}C. 
	As time proceeds (from red to purple), the 0 M suspension (dotted lines) shows only a slight increase in baseline viscosity over all shear stresses, attributed to solvent evaporation.\cite{Brown2012}
	Solvent evaporation increases the suspension packing fraction, resulting in a higher viscosity. 
	Thus, we conclude that there is no significant change in the flow behavior of the 0 M urea suspension over time. 
	
	In contrast, the 6 M urea suspension (solid lines) shows marked differences. 
	First, we see no evaporation for the 6 M urea suspension, which we attribute to the reduction in solvent vapor pressure due to the presence of concentrated urea. 
	Additionally, in the presence of urea, the shear-thinning branch is suppressed, and the thickening onset shifts to higher stresses. 
	Both these effects show no time dependence, so we attribute them to solvated (not sorbed) urea. 
	The presence of urea in solution can screen interparticle attractions that result in shear thinning. 
	Additionally, as we discuss later, the presence of solvated urea could increase the force necessary to push particles into contact, increasing the stresses necessary to achieve direct contact and thus increasing the stress onset of shear thickening. 
	
	In addition to these immediate, time-independent repercussions of the urea presence, we also see a striking time-dependence of the high-shear rheology: as time proceeds, the upper branch of the shear-thickening regime decreases in viscosity. 
	This is particularly important as the high-shear behavior is understood to result directly from the frictional interparticle interactions after hydrodynamic lubrication has broken down. Thus time-dependent changes in interparticle friction will be expected to result selectively in changes to the high-shear rheology. 
	Indeed, the upper Newtonian plateau predicted by the Wyart-Cates model\cite{Wyart2014} decreases as time proceeds. 
	The upper viscosity, $\eta_{\text{max}}$, is plotted relative to the first timepoint (red curves) for both systems in the inset to Figure \ref{Fig3}C. 
	Here we can see that for 0 M urea, the upper viscosity increases over time, consistent with evaporation. 
	In contrast, for 6 M urea, the upper viscosity decreases by nearly 60\% over time, stabilizing around $t=25$ min. 
	The viscosity is determined by the proximity of the suspension packing fraction $\phi$ to the relevant jamming packing fraction, $\phi_J$, as described by the Krieger-Dougherty relation \cite{Krieger1959}
	
	\begin{equation}
	\eta=\eta_0(1-\frac{\phi}{\phi_J})^{-\beta},
	\label{eqn2}
	\end{equation}
	
	where $\eta$ is the suspension viscosity, $\eta_0$ is the pure solvent viscosity, and $\beta$ is a parameter typically close to 2.\cite{Quemada1977,Degiuli2015}
	In the case of the higher-shear, friction-dominated rheology, the relevant jamming packing fraction is the frictional packing fraction $\phi_m$. 
	A lower maximum viscosity indicates that $\phi$ is farther from $\phi_m$. 
	Since $\phi$ is held constant, this indicates that $\phi_m$ must be increasing over time. 
	Since an increasing friction coefficient $\mu$ reduces the frictional jamming point $\phi_m$, we interpret this \textit{increase} in $\phi_m$ to be the result of a \textit{decrease} in the friction coefficient $\mu$. 
	
	It is important to note that we do not anticipate the time dependence of the rheology (\textit{i.e.} an equilibration time of 25 minutes) to match those measured in AFM and NMR studies ($\sim 11$ min). 
	For the AFM and NMR experiments, the concentration of particles is low enough compared to the volume of suspending solvent that the change in the background urea concentration is negligible. 
	In the rheology experiments, where 56\% of the volume of the suspension is occupied by particles, the change in the background urea concentration is no longer negligible and thus the sorption rate will slow down as the urea concentration decreases with time.
	
	\begin{figure}
		\centering
		\includegraphics[width=0.45\textwidth]{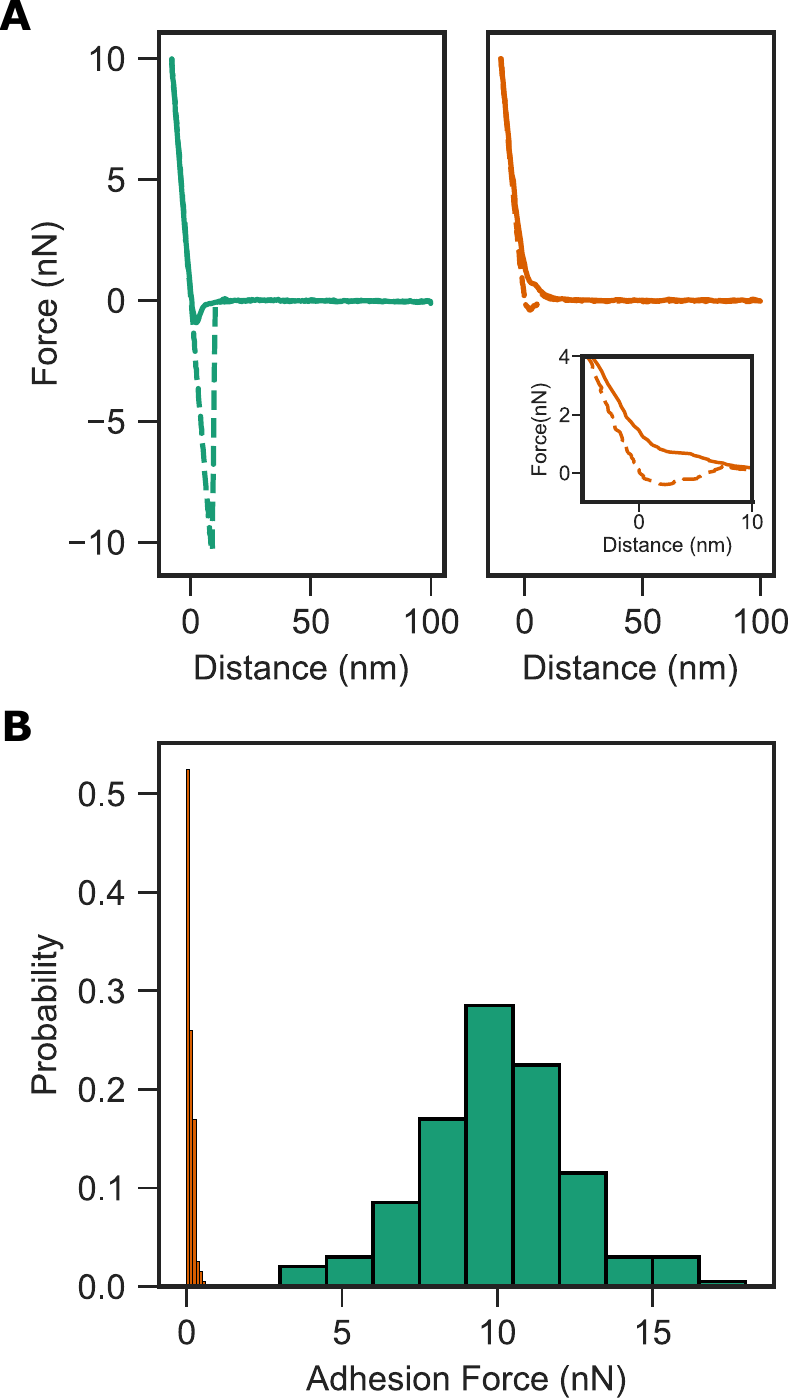}
		\caption{A. Approach (solid line) and retraction (dashed line) of the force-distance curves after 1 hour equilibration in 0 M urea (left) and 6 M urea (right) solutions. Inset: Zoomed-in view to visualize the repulsive shoulder at small approach distances in 6 M urea. B. Adhesion forces measured after 1 hour in 69\% (v/v) glycerol/water at 0 M (teal) and 6 M (orange) urea concentration. The bin width is 1.5 nN for 0 M and 0.1 nN for 6 M.}
		\label{Fig4}
	\end{figure}
	
	To characterize the role of urea in modifying the interparticle adhesion directly, we conduct AFM experiments to measure the adhesion between the PMMA/ITA particle and the carboxylate-functionalized substrate by measuring the pull-off force in a force-vs-distance curve. The force-vs-distance curves are recorded after 1 hour equilibration in 69\% (v/v) glycerol/water at 0 M and 6 M urea concentration . At each urea concentration, 200 curves are recorded across a 20 $\mu$m $\times$ 20 $\mu$m area with the approach and retraction speed set at 500 nm/s.
	
	Representative force-distance curves in 0 M and 6 M urea are shown in Figure \ref{Fig4}A. The 0 M urea retraction curve shows the adhesion forces between the PMMA/ITA particle and the carboxylate-functionalized substrate. This adhesion is due to the hydrogen bonding between the two surfaces.\cite{Noy_Chemical_1995, Sinniah_Solvent_1996} However, the 6 M urea retraction curve shows a massive reduction in the adhesion force. This demonstrates that the presence of urea can effectively deactivate the hydrogen bonding between the surfaces. Additionally, the 6 M urea approach curve shows a small repulsive shoulder when pushing the two surfaces into contact. We attribute this repulsive shoulder to the presence of solvated urea that needs to be squeezed out from the region of contact. Figure \ref{Fig4}B quantitatively compares the adhesion in these systems, showing that an average adhesion of 10.0 $\pm$ 2.5 nN for 0 M urea is reduced to 0.1 nN (95\% confidence interval [0.09, 0.12]) for 6 M urea. 
	
	
	From these results, a picture emerges in which PMMA/ITA particles in suspension experience hydrogen bonding-induced adhesion and enhanced friction. Urea can sorb to the particle surface and reduce this interparticle friction and adhesion. 
	The kinetics associated with equilibration of the sorption process result in time-dependent decreases in the interparticle friction and adhesion, which induce a time-dependent decrease in the upper Newtonian viscosity. 
	These results allow us to pinpoint urea sorption as the driving factor in the reduction of interparticle friction and adhesion, and in turn the shifted SJ regime reported in prior work.\cite{James2018}
	We emphasize the reversible nature of this adhesion generated by particle-particle hydrogen bonds, since aggregation and shear thinning would overwhelm the rheological response and result in a yielding-to-jamming transition, as recently predicted by numerical simulations \cite{Singh_2018}.

	In conclusion, we have presented direct evidence that chemical processes at the particle surfaces and their kinetics influence the interparticle friction and adhesion that drive the macroscopic flow behavior in dense suspensions, specifically shear jamming. 
	The finding that large, stress-activated friction and adhesion can work in concert  introduces a new aspect thatso far has not been taken into account in the literature, although recent work has started to look into different kinds of constraints that contribute to the general mechanisms responsible for DST and SJ\cite{Singh_2018, Guy_2018}. 
	This points to a need for more detailed modeling to map out how adhesion modifies the state diagram that delineates DST and SJ regimes as a function of packing fraction and applied shear stress. 
	Our results also open up new avenues for controlling shear jamming behavior via tuning the particle surface chemistry, in particular developing responsive systems where friction and adhesion can be engineered on demand.

	\begin{acknowledgement}
		The authors thank Shivaprakash Narve Ramakrishna, Endao Han, and Abhinendra Singh for helpful discussions, and Grayson Jackson and Dmitriy S. Dolzhnikov for assistance with the NMR measurements. This work was supported through the Swiss National Science Foundation Grant $PP00P2-172913/1$ and the ETH Research Grant $ETH-49 16-1$. Additional support is acknowledged from the Army Research Office through Grant W911NF-16-1-0078, and from the Chicago Materials Research Center/MRSEC, which is funded by the National Science Foundation through Grant NSF-DMR 1420709. 
	\end{acknowledgement}
	
	\bibliographystyle{unsrt}
	\bibliography{references}
\end{document}